\documentstyle[emulateapj,apjfonts,epsf]{article}

\newcommand{\ud}{\mathrm{d}}

\def\url#1{{\ttfamily\def\/{/\discretionary{}{}{}}#1}}
\def\simless{\mathbin{\lower 3pt\hbox
   {$\rlap{\raise 5pt\hbox{$\char'074$}}\mathchar"7218$}}} 
\def\simgreat{\mathbin{\lower 3pt\hbox
   {$\rlap{\raise 5pt\hbox{$\char'076$}}\mathchar"7218$}}} 

\def\gcm3{{\rm g\,\, cm^{-3}}}

\def\pc3{{\rm pc}^{-3}}

\begin{document}

\title{The contribution of primordial binaries to the blue straggler population in 47~Tucanae}

\author{Michela Mapelli\altaffilmark{1}, Steinn Sigurdsson\altaffilmark{2}, Monica
Colpi\altaffilmark{3}, Francesco R. Ferraro\altaffilmark{4}, Andrea Possenti\altaffilmark{5},
Robert~T. Rood\altaffilmark{6}, Alison~Sills\altaffilmark{7}
\& Giacomo Beccari\altaffilmark{4}
}

\altaffiltext{1}{S.I.S.S.A., Via Beirut 2 - 4, I-34014 Trieste, Italy; {\tt mapelli@sissa.it}}
\altaffiltext{2}{Department of Astronomy and Astrophysics, The Pennsylvania State
University, 525 Davey Lab, University Park, PA~16802; {\tt steinn@astro.psu.edu}}  
\altaffiltext{3}{Dipartimento di Fisica G. Occhialini, Universit\`a di
Milano Bicocca, Piazza della Scienza 3. I-20126 Milano, Italy; {\tt monica.colpi@mib.infn.it}}
\altaffiltext{4}{Dipartimento di Astronomia, Universit\`a
di Bologna, via Ranzani 1, 
I--40126 Bologna, Italy; {\tt ferraro@bo.astro.it}} 
\altaffiltext{5}{Osservatorio Astronomico di Cagliari, Cagliari, Italy; {\tt possenti@ca.astro.it}}
\altaffiltext{6}{Astronomy Dept., University of Virginia, Charlottesville,  
VA 22903-0818, USA; {\tt rtr@virginia.edu}}
\altaffiltext{7}{Department of Physics and Astronomy, McMaster University, 1280 Main Street West,
Hamilton, ON, L8S 4M1, Canada; {\tt asills@mcmaster.ca}}

\begin{abstract}

The recent observation (Ferraro et al. 2003b) of the blue straggler 
population in 47 Tucanae  gives the first 
detailed characterization of their spatial distribution
in the cluster over its entire volume. 
Relative to the light distribution,  
blue stragglers  appear to be overabundant in the core 
and at large radii. The observed 
surface density profile shows a central peak, 
a zone of avoidance and a rise beyond twenty core radii.
In light of these findings we explored the 
evolution of blue stragglers mimicking their dynamics in a  
multi-mass King model for 47 Tucanae.
We find that the observed spatial 
distribution  can not be explained within a purely collisional
scenario in which  blue stragglers
are generated exclusively in the core through direct mergers.
An excellent fit is obtained if we require  that a sizable fraction of 
blue stragglers is generated in the peripheral regions of 
the cluster inside primordial binaries that evolve in isolation  
experiencing mass-transfer.

\end{abstract}

\section{Introduction}

Blue Stragglers (BSs), first discovered by Sandage in
the globular cluster M3,  are stars lying above and blue-ward of the
turn-off point in a cluster color-magnitude diagram.  In recent
years the high angular resolution and the UV imaging capabilities of
the {\it Hubble Space Telescope} (HST) have made possible the search 
of BSs even in the cores of highly concentrated globular
clusters (GCs). Observations indicate that BSs in GCs are
preferentially found in the core (Ferraro et al. 1999a); but in at least
two GCs, M3 (Ferraro et al. 1993; Ferraro et al. 1997) and M55 (Zaggia
et al. 1997) BSs
are seen also in the external region of the GC.   
Ferraro et al. (2003b) have recently found that the 
radial distribution of BSs in 47~Tuc also
appears bimodal, i.e. highly peaked in the core, decreasing at intermediate 
radii and rising  again at larger radii.
A long standing problem is the formation mechanism of
BSs. Many scenarios have been proposed to explain the
BS origin 
(Fusi Pecci et al. 1992; Bailyn 1995; Bailyn 
and Pinsonneault 1995; 
Procter Sills, Bailyn \& Demarque 1995; Sills \&{} Bailyn 1999; Sills et al. 
2000; Hurley et al. 2001) and two seem to be 
the most likely ones.
The first, i.e., the {\it collisional scenario}, indicates that BSs
are the end-product of a prompt merger between two main sequence stars in 
a direct collision that involves a
(resonant) three or four body encounter  
(Davies, Benz \& Hills 1994; Lombardi et al. 2002); these 
BSs acquire kicks generated by
dynamical recoil.  The second, or 
{\it mass-transfer scenario},
suggests that BSs are generated in  primordial
binaries (hereafter PB) that evolve mainly in 
isolation or harden gently by long-distance gravitational encounters
until they reach contact, leading to (unstable) mass-transfer and final
coalescence 
(Carney et al. 2001).
In both these mechanisms 
BSs are formed with a mass exceeding the turn-off mass of the cluster
and can stay on the main sequence through the mixing of the
hydrogen-rich surface layers of its two progenitor stars.
These two scenarios do not necessarily exclude each other and may coexist  
(Leonard 1989; Fusi Pecci et al. 1992; 
Bailyn \& Pinsonneault 1995;
Ferraro et al. 1997; Sills \& Bailyn 1999; Hurley et al. 2001). 
Indeed, bimodal BS radial 
distributions seem to invite the invocation of two mechanisms: the BSs
in the core are principally created by star collisions, while external
BSs are formed from mass-transfer in PBs. Still
Sigurdsson, Davies \& Bolte (1994) suggested that the bimodal population in
M3 might be entirely explained within a collisional model where external BSs
are formed in the core and ejected into the outer regions by recoil.
However, further studies of the BS luminosity function 
and their comparison with theoretical models (Bailyn
\& Pinsonneault 1995; Sills \& Bailyn 1999; Sills et al. 2000; Ferraro
et al. 2003b), highlight the difficulty, if not the impossibility, to
obtain a good fit of the observations assuming that all the BSs are
formed dynamically in the core.
In this paper we study, in the light of the most recent data (Ferraro
et al. 2003b), the case of 47~Tuc, comparing the observed  bimodal
distribution of BSs with a series of simulations carried on using a new
version of the dynamical code described in Sigurdsson \& Phinney~(1995).

\section{Description of the dynamical code}
The code simulates a fixed cluster background with a 
multi-mass King density profile (we choose 10 classes of mass, see Table 1
and section 2 of Sigurdsson \& Phinney 1995) obtained 
imposing a stellar velocity dispersion 
(for a mono-mass cluster) $\sigma{}=11.5\,{} \textrm{km}\,{} \textrm{s}^{-1}$
and a central density $\rho_0=1.26\times{}10^5\,{}M_{\odot{}}\,{}
\textrm{pc}^{-3}$ (Pryor \& Meylan 1993).
The observed
profile has been obtained using the data-set published by Ferraro et
al. (2003b) and the procedure described in Ferraro et al.
(2003c).  Additional HST Archive data in the $V$ band have been used
to combine the HST and the ground based dataset.  All the stars
brighter than the cluster Main Sequence Turn-Off ($V<18$) have been
used to compute the projected density profile.  Fig.~1 shows the
comparison between observed and modeled projected density
profile. This model has a concentration $c=1.95$, and a core radius
$r_c=21\arcsec$.  A typical uncertainty of $\pm 2\arcsec$ can be assumed in
the $r_c$ determination.  Note that the value assumed for 47 Tuc is
consistent with that ($r_c=23\arcsec \pm 2\arcsec$)  obtained
by Howell, Guhathakurta \& Gilliland (2000).
Assuming a distance of 4.6 kpc 
(Ferraro et al. 1999b), the core is $r_c=0.42$ pc.
The dimensionless central potential is $W_0=12$
($W_0\equiv{}\Psi{}(0)/\langle{}\sigma{}\rangle{}^2$, where
$\langle{}\sigma{}\rangle{}$ is the mean core dispersion velocity, 
$\Psi{}(0)\equiv{}\Phi{}(r_t)-\Phi{}(0)$, with $\Phi{}(r)$  the
gravitational potential and $r_t$  the tidal radius (Sigurdsson
\& Phinney 1995).

{ 
\vskip 0.2truecm \epsfxsize=9.truecm \epsfysize=9.truecm
\epsfbox{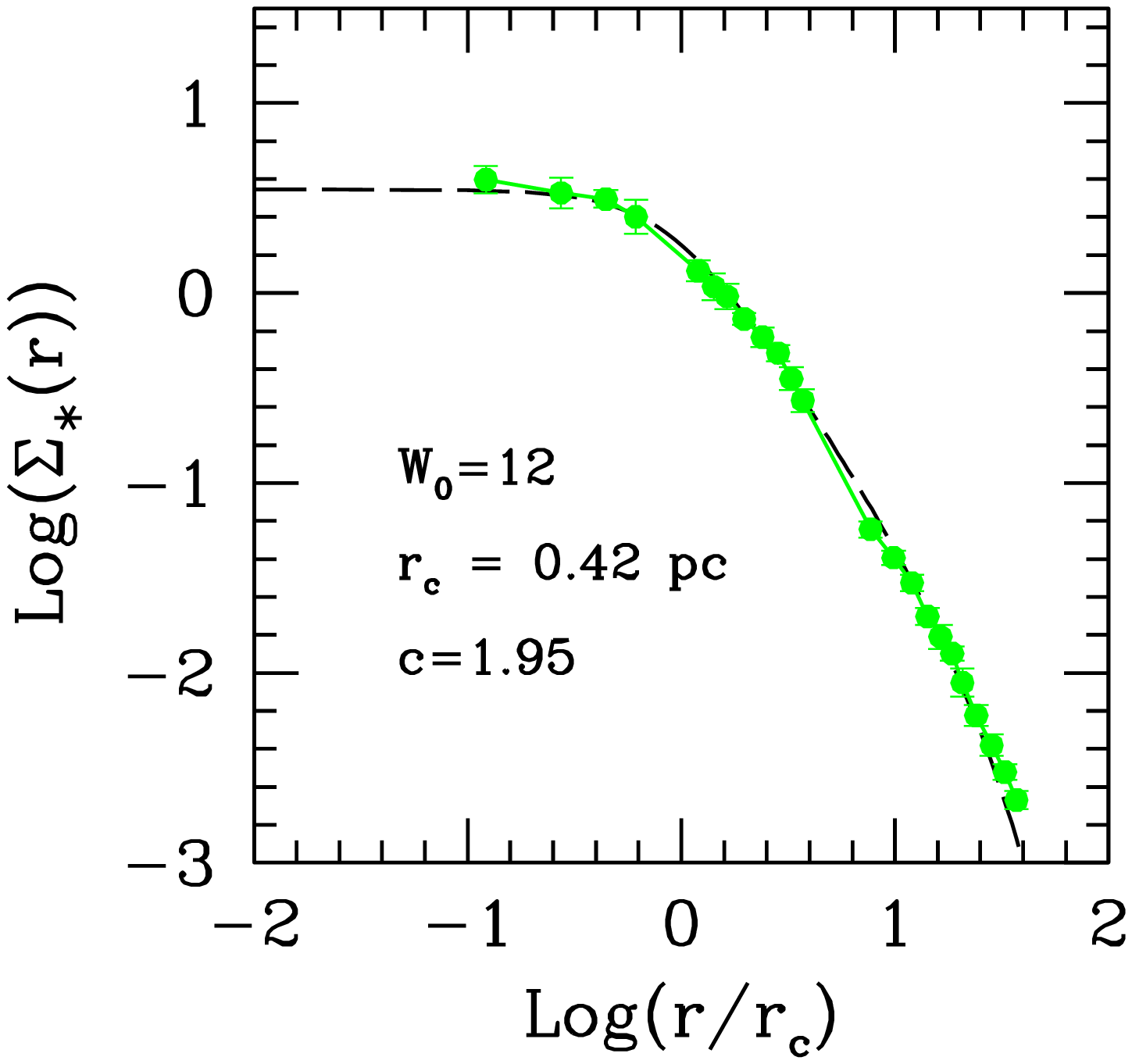}
\label{fig1}
\footnotesize 
{
Fig. 1: Comparison between the observed 
surface density profile of 47~Tuc 
(solid 
line and 
full circles) and the adopted King model (dashed line). 
}

}
\vskip 0.2truecm


On this background we evolve the dynamics of BSs.  
We assume that collisional BSs are generated exclusively  in the innermost
region, at radii less than $n\,{}r_c$, with $n\leq{}=0.1,~0.5$ and $0.8,$
where the density is high to guarantee a high collision rate 
(Pooley et al. 2003).
Internal BSs  generated  
in PBs have been explored as last case and may
be considered as extreme since 
dynamical interactions in dense cluster cores tend to destroy binaries and to 
alter those that remain through exchanges
(Sigurdsson \& Phinney 1993; Ivanova et al. 2004).
External BSs formed in PBs are generated outside the core  with
initial positions distributed in several radial intervals, between 15
and 80 $r_c$.
All initial positions are randomly generated following
a flat probability distribution (see Table 1) according to the 
fact that the number of stars in a King model scales as 
$\ud{}N=n(r)\,{}\ud{}V\propto{}r^{-2}\,{}\pi{}r^{2}\ud{}r\propto{}\ud{}r$.
This is the key difference between our simulations and those of 
Sigurdsson et al. (1995), who generated BSs only in the central region 
(below 0.8 $r_c$).
BS velocities are randomly generated following the distribution
illustrated in section 3 of Sigurdsson \& Phinney 1995
(eq. 3.3). In addition, we assign a natal kick to those BSs formed 
collisionally 
in the core: kick velocities fall in between 1 and 6 $\sigma{}$, and their 
distribution is flat or single valued.
The masses of BSs are generated between 1.2 and 2 $M_{\odot{}}$,
following indications from Ferraro et al. (1997) and Gilliland et al. (1998). 
Every single BS is 
evolved for a time $t_i=t_{\rm last}\,{}*\,{}rand$, where $rand$ is a
random number uniformly generated in [0,1] and $t_{\rm last}$ is the
lifetime assumed for a BS (we have performed sets of runs with
various $t_{\rm last}$, between 1 and 5 Gyr).
Once generated, the BS drifts in the cluster background under the
action of the cluster potential, dynamical friction and distant
encounters (eq. [3.4] of Sigurdsson and Phinney 1995).
Near encounters are uninfluent.

\section {Results}
We performed more than 70,000 runs testing various scenarios.
We first traced the BS evolution in 
the collisional hypothesis (Sigurdsson et al. 1994). We
generated BSs at radii less than 0.8 $r_c$ with recoil velocities in
various ranges. We observed that for recoil velocities greater than 3.5
$\sigma{}$ BSs are expelled from the cluster, while for
recoil velocities lower than 3 $\sigma{}$ dynamical friction drags all
of the BSs into the core in $\sim 10^8$ years.
The final simulated distribution, shown in Fig.~2a,
is peaked in the center, but does
not reproduce the observed rise in the outer region of the GC. 
Thus, no collisional BSs ejected from the core by recoil contribute to
the external population.
The best representation of the data is obtained 
generating  25\% of the BSs 
from mass-transfer in PBs between 30 and 60 $r_c$ with no
kick, and 75\% from collisions inside
0.5 $r_c$ with a natal kick of 1 $\sigma{}$ (see Fig.~2b).
{ 
\vskip 0.2truecm \epsfxsize=8.truecm \epsfysize=8.truecm
\epsfbox{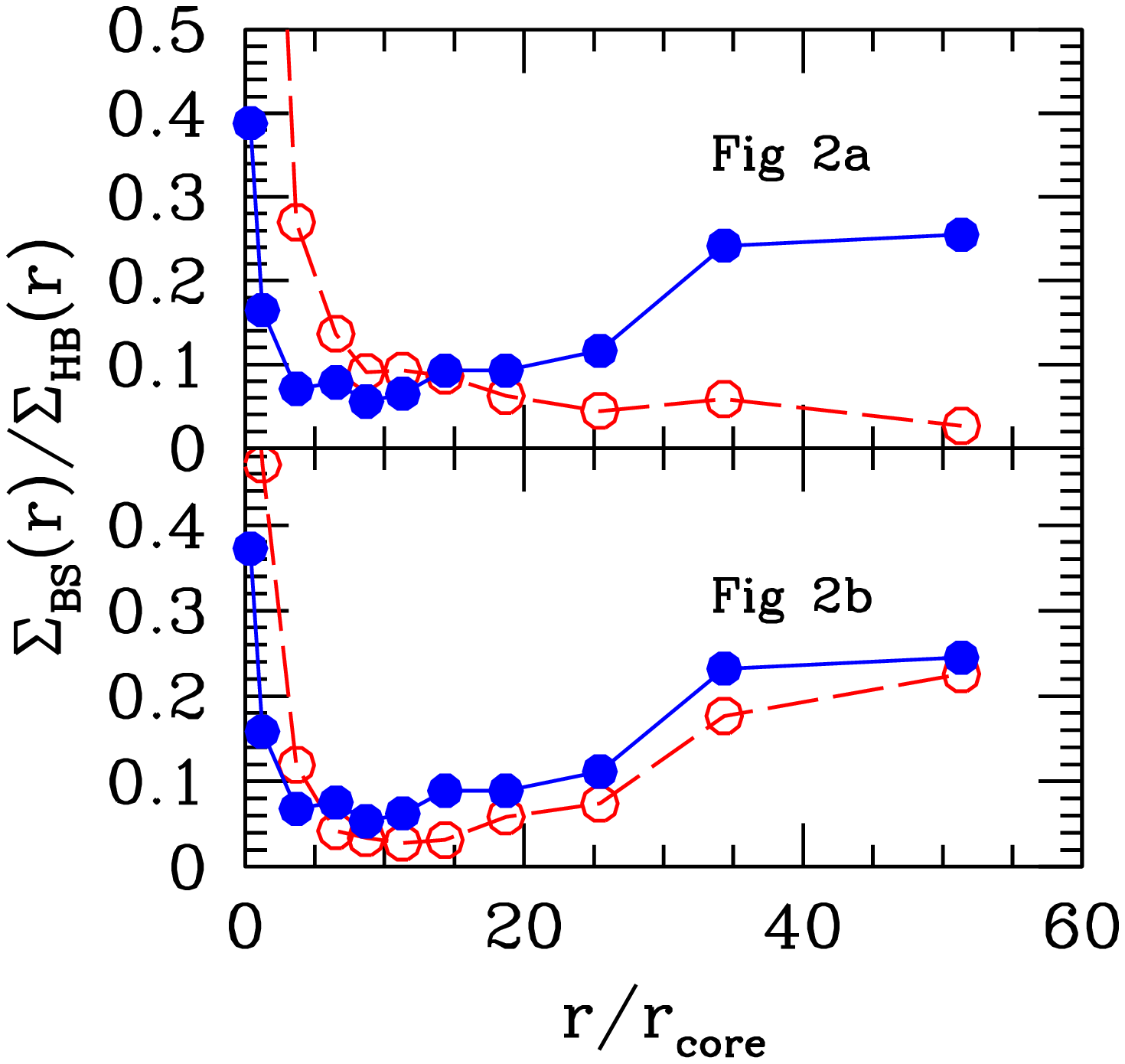}
\label{fig2}
\footnotesize { 
Fig 2: Comparison between the observed BS surface
density (solid line with filled circles) and the simulated BS surface
density (dashed line with empty circles), in the case of BS formation
by collisions only with constant kick = 3.5 $\sigma$ (Fig. 2a), and in the 
case of BS formation both by collisions (with constant kick = 1 $\sigma$) and 
primordial binary mergers with BS lifetime of 1.4 Gyr 
(Fig. 2b). Both the distributions are normalized with respect
to the number of Horizontal Branch stars (HB) cataloged by Ferraro et
al. (2003b). 
}

}
\vskip 0.2truecm

Another key parameter to understand BS evolution is their
lifetime. We performed simulations with various lifetimes $t_{\rm
last}$ (1, 2, 3, 4, 5 Gyr).  We integrated the evolution of every
single BS for a time homogeneously distributed between 0 and $t_{\rm
last}$, assuming that BSs are uniformly generated over the
cluster lifetime, without peaks of formation. In absence of different
indications this seems to be a reasonable assumption (Sigurdsson et
al. 1994). We obtained the best representation for $t_{\rm
last}=1.4-1.5$ Gyr. 
For $t_{\rm last}< 1.4$ Gyr the number of BSs
out of 60 $r_c$ is at least twice the observed value.
For $t_{\rm last}> 1.4$ Gyr, dynamical friction reduces the
amplitude of the  external
rise in BSs and tends to create a secondary ``hump'' of
BSs, which drifts toward the core (Fig. 3). 
At 5 Gyr this hump peaks
at 15 $r_c$, just inside the observed depletion region.
We performed also a set of runs where we impose that BSs are
all generated solely in PBs,  between 
0 and 60 $r_c$ (imposing that 48\% of BSs are generated below 1 $r_c$ 
and 52\% above 1 $r_c$) with no kick. 
Fixing $t_{last}$=1 Gyr, we found a BS 
distribution close to
the observed but with a weaker rise above 30 $r_c$.
{ 
\vskip 0.2truecm \epsfxsize=8.truecm \epsfysize=8.truecm
\epsfbox{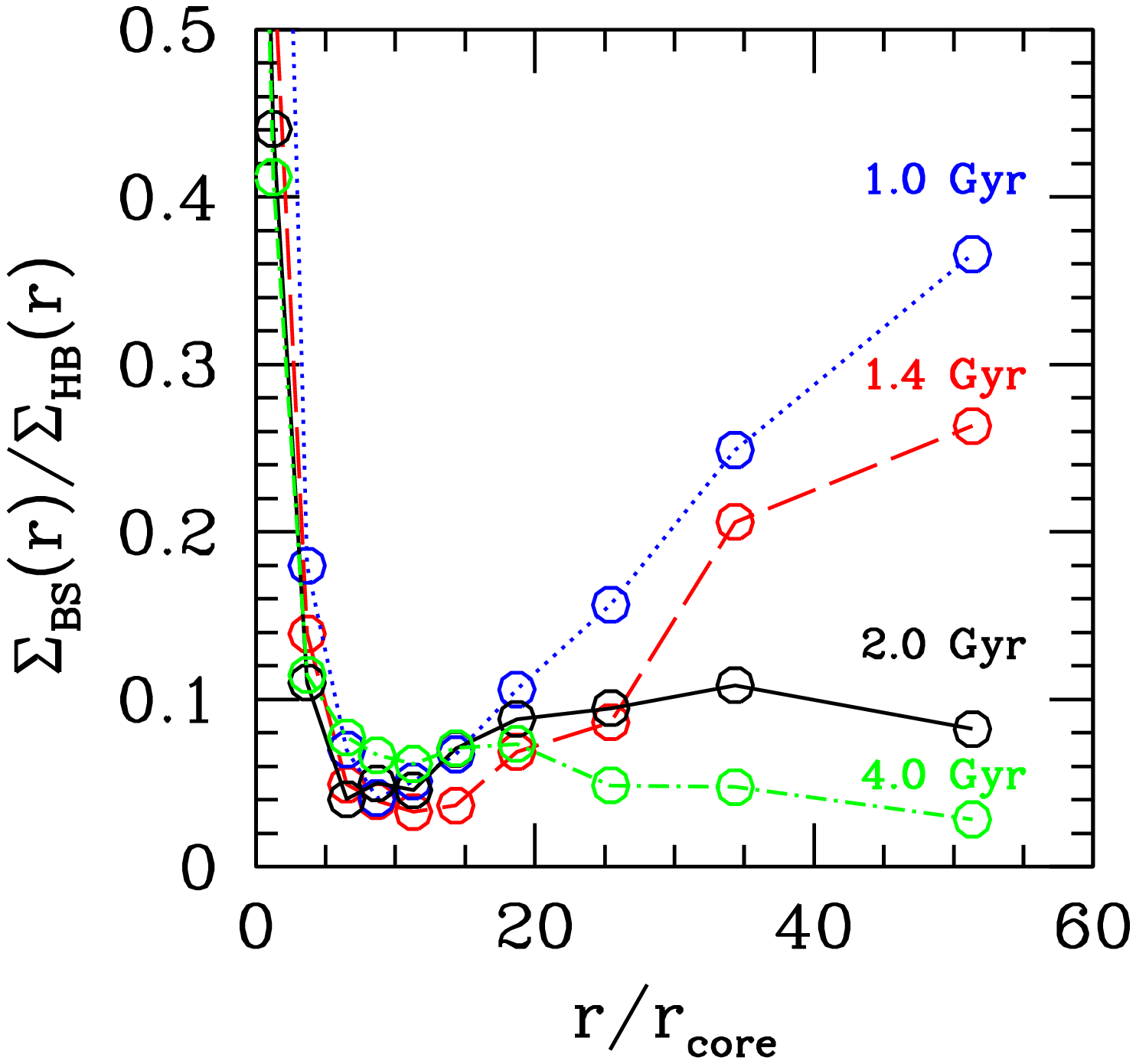}
\label{fig3}
\footnotesize { Fig. 3: Simulated BS surface density at 1 Gyr (dotted line), 1.4 Gyr (dashed 
line), 2 Gyr (solid line) and
4 Gyr (dashed-dotted line) for the same range of initial positions (30-60 
$r_c$), 
initial velocities and masses ($1.2$--$1.8\,{}M_{\odot{}}$). 
}

}
\vskip 0.2truecm


\section {The zone of avoidance}

Our simulations show that all the external BSs are formed by
PBs beyond 30 $r_c$ in order to reproduce 
the spatial distribution with its zone of avoidance. 
Thus we may ask if the external regions of 47~Tuc host a number
of binaries sufficient to produce 
all the observed peripheral BSs and
if these
massive binaries survive in these  regions for  $\sim 12$ Gyr
(approximately the lifetime of the cluster) without drifting toward the
center under the action of dynamical friction.
Unfortunately, there is no direct observational guidance to answer
this question.  The only
study of binaries in 47~Tuc is that of Albrow et al. (2001) who
searched for binaries only in the central regions of 47~Tuc. They found a
binary frequency of $14\%\pm4\%$ in the core of the cluster. However,
the binary fraction in the core, where binaries are created 
or destroyed, or modified in their
properties by exchange interactions, does not provide a direct
measure of the number 
of all truly PBs over the whole cluster. Recent studies by Ivanova et al. (2004)
indicate that the bulk of PBs remain in the outskirts of the GC.  
Searches of PBs  (Rubenstein \& Bailyn 1997
for NGC~6752, and Bellazzini et al. 2002 for NGC~288) are still 
limited to  cluster cores.
Some indication may come from
our simulated cluster model that can provide us the mass
of the cluster out of a certain radius. We found that the mass beyond
30 $r_c$ is $6.4\times{}10^5\,{}M_{\odot{}}$.  We can only guess what
fraction of that mass is in binaries, but an assumption that 10\% of
the mass was originally in binaries is not unreasonable. This gives
$6.4\times{}10^4\,{}M_{\odot{}}$ in binaries or at least
3--$6\times{}10^4$ binaries beyond 30 $r_c$. 
Hurley, Tout \& Pols (2002) estimate a probability of
5$\times{}10^{-4}$  for a binary to generate a BS at 12 Gyr  
(see model F). This implies that our estimated 3--$6\times{}10^4$ PBs 
beyond 30 $r_c$ can produce 15-30 BSs - close enough 
to the 25 required outside  20 $r_c$ (Ferraro et al. 2003b).
Recently, Ivanova et al. (2004) indicate that a
sizable number of binaries survive in the halo ($65\%$) and only a few
percent (5\%) in the core of a simulated 47~Tuc like
GC, assuming an initial binary fraction of 100$\%.$   
If these figures apply, about 100-200 BSs should reside in the halo of
47~Tuc which are not seen; This could be reconciled only if
most of them are faint such to be confused with normal stars.
However, observational studies indicate
that the binary fraction  
tends to be lower in the halo than 
in the core (see for example Rubenstein \& Bailyn 1997).

Two questions now remain: what mechanisms produce
the zone of avoidance~? Can PBs survive 
in the external regions for a time comparable to
the age of the cluster without drifting toward the core
by dynamical friction ? 
We have calculated the
distance from the GC center out of which the dynamical
friction timescale $t_{\rm {DF}}$ 
is longer than the typical age of a PB of 47~Tuc
($\sim{}12$ Gyr). For a BS moving on a circular orbit (Binney \& Tremaine 1987)
\begin{equation} \label{eq:eq1}
t_{\rm {DF}}=\frac{3}{4\ln(\Lambda{})G^2(2\pi{})^{1/2}}\,{}\frac{\sigma{}^3}{M\,{}\rho(r)}
\nonumber
\end{equation}
where $\ln(\Lambda)\sim{}10$ is the Coulomb logarithm, $\sigma{}$ 
the line of sight mean stellar dispersion velocity
($=11.5\,{}\textrm{km}\,{}\textrm{s}^{-1}$), $M$  the mass of a
typical BS ($=1.5\,{}M_{\odot{}}$) and $\rho{}(r)$  the local mass
density at  distance $r$ from the GC
center. We found that the density at
which $t_{\rm {DF}}=12$ Gyr is
$\rho{}(r)\sim{}120\,{}M_{\odot{}}\,{}\textrm{pc}^{-3}$. 
This corresponds to a distance of $\sim{}9\,{}r_c$, as inferred from our King 
model, which is near the position 
of the zone of avoidance. 
In the case of eccentric orbits $t_{\rm {DF}}$  can
be significantly shorter (Colpi, Mayer, \& Governato 1999) and this
opens the possibility that PBs accrete to 
the core from outer distances. By
running simulations that trace the orbital evolution of PBs in the 
GC potential   
with semi-major axis $a\gtrsim{}30\,{}r_c$,  eccentricity
$e\gtrsim{}0.7$ 
and total mass 1.2--$2\,{}M_{\odot{}}$,
we found that in most of our runs PBs
born at $\sim{}60\,{}r_c$ with eccentricity $e=0.7$ have still a
semimajor axis $a\sim{}30\,{}r_c$ after 12 Gyr.
This  suggests that the existence of
a gap in the spatial distribution of the BSs
comes from the interplay between the dynamical friction timescale 
(with its 
spread related to a spread in the orbital parameters)
and 
the characteristic lifetime of the BS.
 
\section{Summary}

The comparison of the observed BS distribution in 47~Tuc with the
simulated distributions supports  the hypothesis that internal BSs principally result from
stellar collisions, while external BSs (outside of 20 $r_c$) are
exclusively generated by mass-transfer in PBs. 
The best fit to the observational data of 47~Tuc is obtained 
when a sizable
fraction (25\%) of BSs is generated  from PBs in peripheral regions ([30,60]
$r_c$). The internal BSs contributing up to 75\%
of the observed are all born inside
0.5 $r_c$ with a natal kick of 1 $\sigma{}$, and do not pollute the
external regions.  External BSs contribute little to the core
population.
A scenario in which all BSs
are generated over the entire cluster by mass-transfer in PBs can not be
ruled out. We 
believe however that a blending between
collision induced evolution and internal evolution is at play in the
GC core to explain internal BSs.
Our main finding is the need of a population of external PBs in order
to generate the bimodal distribution of the BSs observed in 47~Tuc. The 
required fraction (10\%) to fit the data does not seem unreasonable.
More accurate counts of BSs 
in GCs with widely different properties 
are about to be collected  from high resolution photometry (Ferraro et al. 2004
in preparation). These observations may shed light into
the nature of BSs and the importance of PBs in GC evolution. 
In light of these upcoming observations, in  
paper II (Mapelli et al. in preparation),  we plan to continue our
analysis with  our simulations 
over a wide sample
of GCs. Theoretical studies using 
N-body (Baumgardt \& Makino 2003) or Monte Carlo techniques (Ivanova et
al. 2004) will eventually become
necessary tool for exploring
the formation and evolution of BSs in GCs.

\section {Acknowledgments}
The financial support of the Agenzia Spaziale Italiana  and of the {\it
 Ministero dell'Istruzione, dell'Universit\`a e della Ricerca} 
is kindly acknowledged.
RTR is partially supported by STScI
grant GO-8709 and NASA LTSA grant NAG 5-6403.







\newpage
\newdimen\minuswidth    
\setbox0=\hbox{$-$}


\begin{deluxetable}{lll}
\scriptsize 
\tablewidth{0pt}
\tabcolsep 0.12truecm
\tabcolsep 0.12truecm
\tablecaption{Probability Distributions of Initial 
Conditions. \label{tbl-1}}
\tablecolumns{3}
\tablehead{
\colhead{Variable}
 & \colhead{Range} 
 & \colhead{Probability Homogeneous in}
}
\startdata
Mass of the background stars$^{a}$ & [0.1, 25] $M_{\odot{}}$ & Salpeter IMF\\
Initial Position$^{b}$ (r) &     ...     & r \\
Initial Velocity (v) & [0, $\infty{}$] & \parbox[t]{2.5in}{velocity distribution in eq. 3.3 of
Sigurdsson \& Phinney (1995)}\\
$\phi{}$$^{d}$   & [0, $2\pi{}$] & $\phi{}$ \\
$\theta{}$$^{d}$ & [0, $\pi{}$]  & $\cos{\theta{}}$\\
BS Mass  & [1.2, 2] $M_{\odot{}}$ & Salpeter IMF$^{c}$ \\  
BS Lifetime (t)     & [0, $t_{\rm last}$] & t \\
\noalign{\vspace{0.05cm}}
\hline
\enddata
\tablenotetext{a}{We consider 10 mass group, with upper limit for the mass 
respectively: 0.157, 0.20, 0.25, 0.31, 0.39, 0.60, 0.70, 0.80, 1.32, 1.57 
$M_{\odot{}}$. \\
$^{\rm b}$ The range of initial positions depends on the single set of 
runs.\\
$^{\rm c}$ The Salpeter IMF is used to random generate each member of the 
binary progenitor of the BSs.\\ 
$^{\rm d}$ Angles used to random generate the components of the initial 
position and of the initial velocity of BSs.}

\end{deluxetable}

\newpage
\begin{deluxetable}{lllllll}
\scriptsize
\tablewidth{0pt}
\tabcolsep 0.12truecm
\tabcolsep 0.12truecm
\tablecaption{Simulated models. \label{tbl-2}}
\tablecolumns{7}
\tablehead{
\colhead{Set of runs:} &
\colhead{Initial positions ($r_c$)} &
\colhead{Lifetime (Gyr)}&
\colhead{BS Mass  ($M_{\odot{}}$)} &
\colhead{BS fraction within 20 $r_c$} & 
\colhead{BS fraction out of 20 $r_c$} & 
\colhead{Fraction of Escaped BSs} 
}
\startdata

CASE A & [0,0.5] \&  [30,60] & $\quad{}$1   & [1.2,1.8] & 0.80 & 0.11 & 0.09 \\
CASE B & [0,0.5] \&  [30,60] & $\quad{}$1.3 & [1.2,1.8] & 0.83 & 0.11 & 0.06 \\
CASE C & [0,0.5] \&  [30,60] & $\quad{}$1.4 & [1.2,1.8] & 0.79 & 0.12 & 0.09 \\
CASE D & [0,0.5] \&  [30,60] & $\quad{}$1.5 & [1.2,1.8] & 0.82 & 0.11 & 0.07 \\
CASE E & [0,0.5] \&  [30,60] & $\quad{}$2   & [1.2,1.8] & 0.86 & 0.09 & 0.05 \\
CASE F & [0,0.5] \&  [30,60] & $\quad{}$3   & [1.2,1.8] & 0.88 & 0.05 & 0.07 \\
CASE G & [0,0.5] \&  [30,60] & $\quad{}$4   & [1.2,1.8] & 0.88 & 0.05 & 0.07 \\
CASE H & [0,0.5] \&  [30,60] & $\quad{}$5   & [1.2,1.8] & 0.90 & 0.03 & 0.07 \\
CASE I & [0,0.5] \&  [20,60] & $\quad{}$2   & [1.2,1.8] & 0.90 & 0.07 & 0.03 \\
CASE L & [0,0.5] \&  [40,60] & $\quad{}$2   & [1.2,1.8] & 0.84 & 0.07 & 0.09 \\
CASE M & [0,0.8]             & $\quad{}$1   & [1.2,1.8] & 0.84 & 0.02 & 0.14 \\
CASE N & [0,0.5] \&  [30,60] & $\quad{}$2   & [1.2]     & 0.81 & 0.10 & 0.09 \\
CASE O & [0,0.5] \&  [30,60] & $\quad{}$2   & [2.0]     & 0.83 & 0.08 & 0.09 \\
CASE P & [0,60]              & $\quad{}$1   & [1.2,1.8] & 0.74 & 0.14 & 0.12 \\
\noalign{\vspace{0.02cm}}
\hline
DATA   & ---                 & ---          &  ---      & 0.75 & 0.25 & ? \\
\noalign{\vspace{0.02cm}}
\hline
\enddata
\end{deluxetable}


\end{document}